\input ./style/arxiv-general.cfg
\documentclass[aoas,MSNbibl,nameyear,seceqn,dvips]{arximspdf}
\makeatletter
   \@ifpackageloaded{graphicx}{}{\usepackage{graphicx}}
\makeatother

\doi{10.1214/15-AOAS819}
\volume{9}
\issue{2}
\pubyear{2015}
\firstpage{926}
\lastpage{949}
\docsubty{FLA}

\makeatletter

\newcommand{\eqref}[1]{(\ref{#1})}

\newproclaim{definition}{Definition}[section]
\newproclaim{example}{Example}[section]
\newtheorem{proposition}{Proposition}[section]

\newproclaim{assumption}{Assumption}[section]

\newproclaim{remark}{Remark}[section]

\newcommand{\diag}{\operatorname{diag}}

\makeatother

\begin{document}
\begin{frontmatter}

\title{Tracking rapid intracellular movements: A~Bayesian~random~set~approach}
\runtitle{TRIM: A Bayesian approach}

\begin{aug}
\author[A]{\fnms{Vasileios} \snm{Maroulas}\corref{}\ead[label=e1]{vmaroula@utk.edu}}
\and
\author[B]{\fnms{Andreas} \snm{Nebenf\"{u}hr}\thanksref{T1}\ead[label=e2]{nebenfuehr@utk.edu}}
\thankstext{T1}{Supported in part by the NSF (NSF-MCB 0822111).}
\runauthor{V. Maroulas and A. Nebenf\"{u}hr}
\affiliation{University of Tennessee}
\address[A]{Department of Mathematics\\
University of Tennessee\\
1403 Circle Dr.\\
Knoxville, Tennessee 37996\\
USA\\
\printead{e1}}
\address[B]{Department of Biochemistry\\
\quad and Cellular and Molecular Biology\\
University of Tennessee\\
1414 Cumberland Avenue \\
Knoxville, Tennessee 37996\\
USA\\
\printead{e2}}
\end{aug}

%
\received{\smonth{9} \syear{2013}}
%
\revised{\smonth{12} \syear{2014}}

%
\begin{abstract}
We focus on the biological problem of tracking organelles as they move
through cells. In the past, most intracellular movements were recorded
manually, however, the results are too incomplete to capture the full
complexity of organelle motions. An automated tracking algorithm
promises to provide a complete analysis of noisy microscopy data. In
this paper, we adopt statistical techniques from a Bayesian random set
point of view. Instead of considering each individual organelle, we
examine a random set whose members are the organelle states and we
establish a Bayesian filtering algorithm involving such set states. The
propagated multi-object densities are approximated using a Gaussian
mixture scheme. Our algorithm is applied to synthetic and experimental data.
\end{abstract}

%
\end{frontmatter}

\section{Introduction}\label{intro}
Most plant cells display a striking phenomenon called ``cytoplasmic
streaming,'' a process that has been recognized since the late 18th
century by \citet{Cor}. During cytoplasmic streaming, most subcellular
organelles move rapidly through the cell, resulting in constant mixing
of the soluble components of the cytoplasm. The function of these
movements is not known, although a potential role in better
distribution of metabolites has been proposed in \citet{ShYo}. The
movements are driven by myosin motor proteins [\citet{Shi}] and appear
to be necessary for normal growth of plant cells and ultimately the
whole plant [\citet{PPA,OTP}]. The molecular mechanisms that connect
the intracellular movements with cell growth are not known [\citet
{MaNe}]. Better understanding of these cellular processes requires the
targeted manipulation of the movements followed by quantitative
assessment of the resulting changes at the subcellular, cellular and
whole-plant levels. Recent results have identified additional
regulatory mechanisms that influence intracellular movements, although
the precise nature of these mechanisms is still unknown [\citet{ViNe}].
This is due, at least in part, to the astounding complexity of these
movements and the technical difficulty of describing them accurately
[\citet{Neb,HTF}].

Recent advances in molecular biology and fluorescence microscopy
imaging have made possible the detailed observation of these
intracellular dynamics and the acquisition of large multidimensional
image data sets [\citet{Dan}]. \citet{PSE} noted that these time-lapse
observations reveal a large number of nearly identical particles that
move with high velocities in close proximity to each other. Combined
with the saltatory, or stop-and-go, nature of their motions, these
features make automated tracking of these movements an extremely
difficult task as discussed in \citet{Neb}. As a result, most previous
analyses have relied on manual tracking of a few individual particles,
for example,\break \citet{Neb,GLP,HTF,LoLe,Col}. A full understanding of the
observations, however, requires accurate tracking of a large number of
bright spots in noisy image sequences, which can be accomplished only
by an automated algorithm that is able to analyze the data completely
[\citet{Dan}]. This complete analysis will require reliable
identification of organelle positions (coordinates) from the bright
spots in fluorescent microscope images taken at different times and the
correct linking of these positions into continuous movement
trajectories over all time points. One benefit of such an algorithm
could be the emergence of recurring patterns such as the recent
discovery, based on manual tracking, that organelles preferentially
pause their motions at microtubules [\citet{HTF}]. Thus, it seems
likely that a comprehensive and accurate tracking algorithm will
unearth additional regulatory events that in turn can be studied experimentally.
Moreover, from a statistical point of view, an automated tracking
algorithm will reduce the bias since manual tracking depends solely on
experts' decision of linking the positions of bright spots at
subsequent time points.

Mathematical and statistical models that require knowledge from
statistics, probability, scientific computing and statistical mechanics
have been developed for reliably tracking multiple objects in space.
There are a great number of studies addressing the problem of tracking
multiple targets in various settings. A partial list of such works is
\citet{DFG,Liu,GMN,GiBe,FBS,ShBl,BlPo,LiCh,MaSt,VVC}, \citeauthor{Mah}
(\citeyear{Mah,Mah3}), \citet{MaMa}.
However, only a small number of multi-object models have been
considered for specific microscopy image data, for example, \citet
{Sma,SNM,SbKo,JLM1}. Movement of subcellular particles in living cells
poses a highly complex problem for automated tracking algorithms. Even
at high magnification, the true position of a particle within a cell
can be only measured to within 50--200 nm due to limitations in optical
resolution, and given the inevitable image noise, it is likely that
some organelles are not detected. Moreover, not only can individual
organelles move independently, they also can change their behavior
rapidly, their paths are not static, and organelles in close proximity
can display strikingly different behaviors [\citet{Col,Neb}].
Commercial automated tracking algorithms such as Perkin--Elmer's
``Volocity'' were sometimes used to gain insights into overall movement
patterns or derive average movement velocities; for example, see \citet
{PPA,APM}. However, these algorithms often introduced mis-assignments
in the tracks [e.g., Figure~3A in \citet{APM}] and, therefore, cannot
be used to obtain an accurate global view of organelle motility.

In general, from a statistical point of view, tracking of multiple
objects is an inherently difficult problem and consists of computing
the best estimate of the objects' trajectories based on noisy
observations. The estimates are propagated by a posterior distribution
which considers organelles' dynamics and combines them with data. The
greater the number of objects that are being tracked, the more
complicated the tracking algorithm becomes. There are several
techniques, for example, Kalman filters and their derivatives, particle
filters, for addressing this problem statistically. The reader may
refer to \citet{GSS,Liu} and the references therein.

A popular approach to tracking is particle filtering. \citet{SDG}
introduced a particle filtering algorithm for the tracking problem
using microtubule dynamics, which overall follow a priori known and
fairly straight paths and can therefore be conveniently modeled. In
general, the particle filter
approach is an importance sampling method which approximates the
posterior distribution by a discrete set of weighted samples
(particles). However, it is often found in practice that most samples'
contribution to the posterior distribution will be negligible.
Therefore, carrying them along does not contribute significantly to
finding an estimate.
Hence, one may resample the particles to create more copies of samples
with significant weights [\citet{GSS}]. However, even with the
resampling step, the particle filter might still need a large number of
samples in order to approximate accurately the target distribution.
Typically, a few samples dominate the weight distribution, while the
rest of the samples are in statistically
insignificant regions [\citet{SBB}]. Thus, some studies [see, e.g.,
\citet{GiBe,MaSt,Wea,KaMa}] have used an additional Markov Chain Monte
Carlo step which helps to move more samples into statistically
significant regions and thus to improve the diversity of samples. This
extra step can improve estimates for multi-target tracking scenarios
[\citet{MaSt,KMS}], but at the price of adding an additional layer of
complexity.

In this manuscript, we attempt to avoid the technical algorithmic steps
which depend on the specific nature of different applications. Instead,
we create an automated statistical tracking algorithm for independently
evolving intracellular movements by considering a pertinent
multi-object statistical framework. This framework adopts a Bayesian
random set filtering technique. The key innovation in our approach is
to conceptually view the
evolving collection of organelles as a single set-valued state and
the collection of the experimental measurements as a single
set-valued observation. A set-valued state contains not only the
position of existing organelles but also the states of new biological
entities which enter the tracking domain. Using Random Finite Set (RFS)
theory and modeling the collection of organelles and their
corresponding experimental measurements as sets result in \emph
{generalizing single-object filtering to a rigorous formulation of
Bayesian multi-object filtering}. Multi-object filtering, similar to
the single-object case, consists of two stages, the prediction stage
using modeled or experimentally derived dynamics, and the update stage
using the observed data. Both these steps involve multi-object
distributions which lead to the multi-object Bayesian filtering
posterior distribution,
\[
f(X|Z_1,\ldots,Z_t) \propto f(Z_{t}|X)f(X|Z_1,
\ldots,Z_{t-1}),
\]
where $X, Z_1,\ldots, Z_{t}$ are appropriate random sets, formally
defined in Section~\ref{RFS}.

The general multi-object Bayes filtering distribution, $f(X|Z_1,\ldots
,Z_t)$, is, however, computationally intractable in most applications
and thus it needs to be approximated. In this paper, we consider a
Gaussian mixture Cardinalized Probability Hypothesis Density (CPHD)
approximation. The CPHD, first introduced by \citet{Mah}, propagates
two estimates, the cardinality distribution of a random set which
yields an estimate of the number of objects per time step, and the
intensity of a random finite set or otherwise the so-called probability
hypothesis density (PHD) [\citet{Mah3}]. The PHD is similar to the
first-moment density or intensity density in point process theory; for
example, see \citet{DaVe}. The PHD first monitors multiple objects as
clusters, and then attempts to resolve individual objects only as the
quality and quantity of data permits. One could also estimate the
number of objects at a given time step using the PHD, however, such an
estimate is unstable when the experimental scene is highly dynamic,
that is, with rapid entry and exit of organelles from the region of
interest. A~Gaussian mixture approximation of the CPHD was introduced
by \citet{VVC} whose algorithmic complexity was of the order $\mathcal
{O}(m^3n)$, where $m$ is the number of data points (acquired positions
of organelles) and $n$ the true number of objects of interest. However,
the cubic dependency on the number of data points is disadvantageous
for our biological framework due to their large number.

In our manuscript, we consider a Gaussian mixture CPHD based on the
experimental fact that data are generated only when organelles are
present in the tracking domain. A~false alarm is generated in signal
detection when a nontarget event exceeds the detection threshold. Our
experiments did not suffer from any false alarm, and thus a pertinent
approximation of the CPHD is established in Propositions \ref{gpred}
and~\ref{gcorr}. The associated algorithmic implementation cost
reduces to the order of $\mathcal{O}(mn)$, that is, the cost is linear
with respect to the number of data and organelles. In brief,
Proposition~\ref{gpred} propagates the predicted cardinality and the
predicted intensity estimate (PHD) of a random finite set which follows
a Gaussian mixture density. Taking into consideration a new random set
of data (positions of organelles), Proposition~\ref{gcorr} updates the
two predictions by considering a Bayesian set formulation. The
posterior PHD follows an appropriate Gaussian mixture whose components
are derived with the aid of Proposition~\ref{gcorr}.

A similar algorithm was analyzed in \citet{MaMa} for the special case
of monitoring two fixed objects that spawn several objects along their
ballistic trajectories. These secondary objects fall under gravity, and
thus they are not of tracking interest. Precisely, a distance criterion
was computed to distinguish the two primary objects from the spawned
ones. When this distance exceeded a certain threshold, the
corresponding objects were declared debris and they were discarded.
This assumption cannot be incorporated herein. Thus, in our framework,
we relax this condition and, moreover, we incorporate several
experimental biophysical features to understand the unknown dynamics of
organelles. For instance, based on the organelles' acceleration data
analysis (see Section~\ref{results}), we discover that the acceleration
follows a normal distribution with mean-zero. Assuming that the mass of
the observed organelles did not change significantly between individual
images (a valid assumption), we are able to deduce interesting results
about the developed biomechanics within a cell.

Section~\ref{RFS} focuses on the methodology that was followed to
establish an automated tracking algorithm for organelle movement data.
Definitions of the Cardinalized Probability Hypothesis Density (CPHD)
and approximation schemes are also presented. Section~\ref{results}
describes the implementation of an appropriate version of the Gaussian
mixture CPHD filter suited for the biological data (synthetic and
experimental). Section~\ref{data} describes the biophysical conditions
under which the experimental data were collected and the process of
manual tracking. Finally, our results are summarized in Section~\ref
{summary} and a discussion for future research and developments is offered.
%
%
\section{Random finite sets and approximations} \label{RFS}
We motivate this section by considering first the problem of tracking
only one object. Suppose that an organelle, whose state is $x'$ at time
$t$, moves following the dynamics below,
%
\begin{equation}
\label{ddyn} x_{t+1}= \phi_t\bigl(x',
u_t\bigr),
\end{equation}
where $u_t$ is a randomly distributed noise and $\phi_t\dvtx \mathbb{R}^N
\times
\mathbb{R}^N \rightarrow\mathbb{R}^N$ is a family of nonlinear,
nonsingular functions. Let $z_{1:t} \doteq\{z_1,z_2,\ldots, z_t\}$
denote the data history up to time $t$ and let $f_{t|t}(x'|z_{1:t})$
represent the posterior probability density function (p.d.f.) at a given
time $t$. Furthermore, consider the posterior predictive p.d.f.,
$f_{t+1|t}(x|z_{1:t})$, which merely yields the probability that an
organelle will move to state $x$ at time $t+1$ given the available data
$z_{1:t}$. Using the Chapman--Kolmogorov equation, the posterior
predictive distribution is given by
%
\begin{equation}
\label{singlepred} {f_{t+1|t}(x|z_{1:t}) = \int f_{t+1|t}
\bigl(x|x'\bigr)f_{t|t}\bigl(x'|z_{1:t}
\bigr) \,dx'},
\end{equation}
where $f_{t+1|t}(x|x')$ is the Markov transition density associated
with the dynamics expressed of equation \eqref{ddyn}. At given time
$t+1$, a new microscopy observation is collected, $z_{t+1} \in\mathbb
{R}^M$. Typically, the dimension of organelle states, $N$, and the
dimension of data, $M$, are not identical, $N \neq M$. For example, the
state of organelles involves their position on the $xy$-plane and the
corresponding velocities, that is, $N=4$, whereas only the positions
($M=2$) are available from the experimental data. The prediction (\ref
{singlepred}) needs to be updated using the datum $z_{t+1}$. The
collected measurement is a function of the true organelle's state
perturbed by noise, that is,
%
\begin{equation}
\label{obs1} z_{t+1}= \eta_{t+1}(x,\xi_{t+1}),
\end{equation}
where $\xi_{t+1}$ is a randomly distributed noise, independent from
$v_t$, and the function $\eta_{t+1}\dvtx \mathbb{R}^N \times\mathbb{R}^M \rightarrow\mathbb{R}^M$ is a
family of nonsingular, nonlinear transformations. Based on the Bayesian
rule, the posterior p.d.f. at a given time $t+1$ is given by
%
\begin{equation}
\label{filtering} f_{t+1|t+1}(x|z_{1:t+1})= \frac{f_{t+1}(z_{t+1}|x)f_{t+1|t}(x|z_{1:t})}{
\int f_{t+1}(z_{t+1}|x)f_{t+1|t}(x|z_{1:t})\,dx},
\end{equation}
where $f_{t+1}(z|x)$ is the likelihood function associated with \eqref
{obs1} and the posterior predictive distribution,
$f_{t+1|t}(x|z_{1:t})$, is defined in \eqref{singlepred}.

\begin{remark}
The widely-known Kalman filter is a special case of the Bayesian
filtering formulation given in equation \eqref{filtering}. Indeed, if
one considered that $\phi_t,  \eta_t$ were linear and $v_t, w_t$ were
normally distributed, then equations \eqref{singlepred} and \eqref
{filtering} would enjoy a closed-form solution which would be the same
as in the Kalman filter.
\end{remark}

On the other hand, our focus is on tracking multiple objects which move
simultaneously. Motivated by the single object tracking framework
described in equations \eqref{singlepred} and \eqref{filtering}, we
consider \emph{a statistical framework which allows us to generalize
the prediction equation \textup{\eqref{singlepred}} and the corresponding update
equation \textup{\eqref{filtering}}, both suitable for tracking one object to
pertinent equations for tracking one set of objects}. We view for the
first time in this biological problem the evolving collection of the
organelles as a \textit{single set-valued state}, $X_t=\{ x_t^1, x_t^2,
\ldots, x_t^{n_t}\} \in
\mathcal{F}(\mathbb{R}^N)$, where $n_t$ represents the number of
objects at time
$t$, and $\mathcal{F}(\mathbb{R}^N)$ is the
collection of all finite subsets of $\mathbb{R}^N$. Similarly, the
collection of experimental microscopy measurements at time $t$ is
viewed as a \textit{single
set-valued observation}, $Z_{t}=\{z_{t}^1, z_{t}^2, \ldots,
z_{t}^{m_{t}}\} \in
\mathcal{F}(\mathbb{R}^M)$, where $m_{t}$ is the
number of generated measurements at time $t$. Based on equation \eqref
{obs1}, each member $z_{t}^i \in Z_{t+1}$ is a noisy perturbation of
the true state $x$ of an organelle $j$ at time $t$, where $i$ is not
necessarily equal to $j$.

Furthermore, the randomness in this multi-object framework is
represented by modeling multi-object
states, $\Delta_t$, and multi-object measurements, $M_t$, as random
finite sets (RFS)
on the single-object state and observation spaces, $\mathbb{R}^N$
and $\mathbb{R}^M$, respectively. The corresponding multi-object
dynamics and
observations are described below.

Given a realization, $X_{t}$, of the RFS, $\Delta_t$, at time $t$, the
multi-object state at time $t+1$ is modeled by the RFS,
%
\begin{equation}
\label{preddynamics} \Delta_{t+1} = \biggl\{ \bigcup
_{x \in X_{t}} S_{t+1|t}(x) \biggr\} \cup B_{t+1},
\end{equation}
where $S_{t+1|t}$ is the RFS representing the objects which survive
with probability $p_{S,t+1|t}(x)$,
from the previous time $t$,
and $B_t$ is the RFS which represents the objects which enter
the scene at time $t+1$ (``newborn'' organelles). Hence, the RFS,
$\Delta_{t+1}$, includes all information of set dynamics,
such as the number of objects that vary over time and an individual organelle's
motion [see equation \eqref{ddyn}] and birth/death. Now, given a
realization $X_{t+1}$ of
$\Delta_{t+1}$ at time $t+1$, the multi-object measurements are
modeled via the following RFS,
%
\begin{equation}
\label{corrdynamics} M_{t+1} = \bigcup_{x \in X_t}
\Theta_{t+1}(x),
\end{equation}
where $\Theta_{t+1}(x)$ is the RFS
of measurements generated by the object $x \in X_t$. The RFS
$M_{t+1}$ encapsulates all characteristics of the measurements from the
microscopy image, for example, measurement noise.

Next, let $f_{t|t}(X'|Z_{1:t})$ denote the multi-object posterior
density at a given time step $t$ conditioned on the observation sets, $Z_{1:t}
\doteq\{ Z_1, Z_2, \ldots, Z_t\}$. The
multi-object Bayes filter propagates the multi-object filtering
distribution via the following recursion:
%
\begin{eqnarray}
f_{t+1|t}(X|Z_{1:t}) &=& \int f_{t+1|t}
\bigl(X|X'\bigr)f_{t+1|t}\bigl(X'|Z_{1:t}
\bigr) \delta X', \label{multimotion}
\\
f_{t+1|t+1}(X|Z_{1:t+1}) &=& \frac{f_{t+1}(Z_{t+1}|X)f_{t+1|t}(X|Z_{1:t})}{
\int f_{t+1}(Z_{t+1}|X)f_{t+1|t}(X|Z_{1:t}) \delta X}, \label{multifiltering}
\end{eqnarray}
where $\int\delta X$ is the set integral [see, e.g.,
\citet{GMN}, Definition~10], $f_{t+1|t}(X|X')$ is the multi-object transition
density associated with the dynamics given in equation (\ref
{preddynamics}), and
$f_{t+1}(Z_{t+1}|X)$ is the multi-object likelihood obtained by equation
(\ref{corrdynamics}). One may show that densities and likelihoods
expressed in equations \eqref{multimotion} and \eqref{multifiltering}
are well defined using techniques of finite set statistics (FISST) and
extending the concept of the Radon--Nikodym derivative [\citet{GMN}, Chapter
II.5].

\begin{remark}
One may compare the analogy between equations \eqref{multimotion},
\eqref{multifiltering} and equations \eqref{singlepred}, \eqref
{filtering}, respectively. Therefore, our statistical framework
generalizes the problem from tracking a single object to tracking a
single set.
\end{remark}

However, the multi-object filter described in equations
(\ref{multimotion}) and (\ref{multifiltering}) is intractable in most
applications
and the Cardinalized Probability Hypothesis Density (CPHD)
approximation is considered. The CPHD produces estimates on the number
of organelles and their states. A formal definition is below.

\begin{definition} \label{CPHD}
The CPHD filter recursively propagates the posterior cardinality
distribution $
p_{t|t}(n|Z_{1:t})$ on object-number $n$ and
the intensity function or Probability Hypothesis Density (PHD)
$D_{t|t}(x|Z_{1:t})$.
Given any region $S\subseteq\mathbb{R}^{N}$, the expected number of
objects in $S$ is derived by the integral $\int_{S}D_{t|t}(x|Z_{1:t})\,dx$. If $S=\mathbb{R}^{N}$, then
$
N_{t|t}=\int D_{t|t}(x|Z_{1:t})\,dx
$
is the total expected number of objects in the scene.
\end{definition}

The CPHD filter produces stable (low-variance) estimates of object
number, as well as better estimates of the states of individual
objects [\citet{Mah,VVC}]. This gain in performance is achieved with
increased computational cost. For instance, \citet{VVC} implemented a
Gaussian mixture CPHD whose algorithmic cost was of the order $\mathcal
{O}(m^3n)$, where $m$ is the number of data points and $n$ the number
of objects of interest. However, the number of data points is large and
the number of organelles is a priori unknown and varies in time.
Therefore, the alternative Gaussian mixture implementation of \citet
{MaMa} is considered herein which decreases the computational cost to
the order of $\mathcal{O}(mn)$. In fact, our technique is based on the
experimental observation that all data are produced by the organelles
and no false alarms exist. If false alarms were collected, for
instance, due to human intervention, then equations \eqref
{preddynamics} and \eqref{corrdynamics} would need to be suitably formulated.

Before proceeding with the dynamics and Bayesian formulations as
expressed in Propositions \ref{gpred} and~\ref{gcorr},
respectively, we list the assumptions on which our Gaussian mixture
approach to the CPHD is based.

\begin{assumption} \label{gsd}
Consider a realization $X_t=\{ x_t^1, x_t^2, \ldots, x_t^{n_t}\}$ of
the RFS, $\Delta_t$, and the associated data collection $Z_t=\{
z_t^1,z_t^2,\ldots, z_t^{m_t}\}$. The state of each organelle $x_t^i
\in X_t,   i=1, \ldots,n_t $ is normally distributed given by
%
\begin{equation}
x_t|x_{t-1} \sim N(x;F_{t-1}x_{t-1},Q_{t-1}),
\label{gaustrans}
\end{equation}
where $F_{t-1}$ is the state transition matrix and $Q_{t-1}$ is the
process noise covariance. Similarly, each observation $z_t^j
j=1,\ldots,m_t,  j \neq i$, is normally distributed according to
%
\begin{equation}
z_t|x_t \sim N(z;H_{t}x_t,R_{t}),
\label{gausslik}
\end{equation}
where $H_{t}$ is the observation matrix and
$R_{t}$ is the observation noise covariance.
\end{assumption}

\begin{assumption} \label{prob}
The survival probability, $p_{S,t+1|t}(x)$, of an organelle with state
$x$ at time $t$ to be present at time $t+1$ is state independent,
that is, $p_{S,t+1|t}(x) = p_S$. The detection probability,
$p_{D|t+1}(x)$, to collect an observation associated with an organelle
whose state is $x$ at a given time $t$, is state independent, that is,
$p_{D|t+1}(x) = p_D$.
\end{assumption}

\begin{assumption} \label{gbd}
The intensity measure of the birth RFS which encompasses the dynamics
of newborn organelles is a Gaussian mixture of
the form
%
\begin{equation}
\label{bithgauss} b_{t}(x)= \sum_{i=1}^{J_{b,t}}
w_{b,t}^{(i)} N\bigl(x;\mu_{b,t}^{(i)},P_{b,t}^{i}
\bigr),
\end{equation}
where $w_{b,t}^{(i)}, \mu_{b,t}^{(i)}, P_{b,t}^{i}$ are the
weights, means and covariances of the mixture birth intensity and
$J_{b,t}$ is the number of Gaussian components associated with the
newborn organelles at a given time $t$.
\end{assumption}

\begin{remark}
Assumptions \ref{gsd}--\ref{gbd} are crucial for establishing a closed
form for the multi-object densities defined in equations (\ref
{multimotion}) and (\ref{multifiltering}). However, if the
linearity of Assumption~\ref{gsd} is violated, then one could consider
implementing a CPHD filter introduced by \citet{VVC}, which employs a
pertinent approximation of the nonlinearities. However, Assumptions
\ref
{gsd}--\ref{gbd} are satisfied using our experimental data. Further
discussion of this topic is delegated to Section~\ref{results}.
\end{remark}

The propositions below involve the main equations of the Gaussian
mixture implementations of the CPHD filter without considering any
false alarms. For presentation's sake, the time index is suppressed in
the cardinality of the state sets and measurement sets in the
propositions below, that is, $n_t =n$ and $m_{t+1} =m$. The reader
should refer to \citet{MaMa} and the references therein for their proofs.

\begin{proposition}[(Prediction)] \label{gpred}
 Assume that at a given time $t$, the posterior cardinality
distribution, $p_{t|t}(n)$, is given and that the posterior PHD is a
Gaussian mixture of the form $D_{t|t}(x) = \sum_{i=1}^{J_{t}}
w_{t}^{(i)} N(x;\mu_{t}^{(i)},P_{t}^{(i)})$, where $J_t$ is the number
of Gaussian components at $t$. Then the posterior predicted PHD,
$D_{t+1|t}$, is also a Gaussian mixture,
%
\begin{equation}
\label{gpredZFAPHD} D_{t+1|t}(x)= b_t(x) + D_{S,t+1|t}(x),
\end{equation}
where $b_t(x)$ is given in (\ref{bithgauss}) and
$D_{S,t+1|t}(x)=p_S\sum_{i=1}^{J_{t}} w_{t}^{(i)}
N(x;\mu_{S,t+1|t}^{(i)},\break  P_{S,t+1|t}^{(i)})$ is the PHD which arises
from the ``survived'' organelles. The corresponding mean and covariance equal
$\mu_{S,t+1|t}=F_t\mu_{t}$ and $P_{S,t+1|t}=Q_t+F_tP_tF_{t}^T$, respectively.
The posterior predictive cardinality distribution is
%
\begin{equation}
\label{gpredZFAcard} p_{t+1|t}(n|Z_{1:t}) = \sum
_{j=0}^n p_B(n-j)\sum
_{l=j}^\infty \pmatrix{l
\cr
j} p_{S}^j
(1-p_S)^{l-j} p_{t|t}(l),
\end{equation}
where $p_B(\cdot)$ is the cardinality distribution of the RFS
responsible for the organelles' appearance and $p_S$ is the survival
probability of an organelle.
\end{proposition}

We denote the permutations $P_m^n=\frac{n!}{(n-m)!}$ with the
convention that $P_m^n=0$, if $n<m$, and we define $q_D=1-p_D$ the
probability of not detecting an intracellular movement. Furthermore,
assume that at time $t+1$, a new measurement random set, $Z_{t+1}$, is
received with cardinality $|Z_{t+1}|=m$. Then the predicted PHD \eqref
{gpredZFAPHD} and cardinality distribution \eqref{gpredZFAcard} will be
updated according to Proposition~\ref{gcorr}.

\begin{proposition}[(Update)] \label{gcorr}
Suppose that the predicted PHD, $D_{t+1|t}$, and the cardinality
distribution, $p_{t+1|t}(n|Z_{1:t})$, satisfy Proposition~\ref{gpred}.
Then, the posterior PHD, $D_{t+1|t+1}$, at a given time $t+1$ is a
Gaussian mixture, and the corresponding CPHD
update equations are listed below:
%
\begin{eqnarray}
\label{gcorrZFAPHD} D_{t+1|t+1} &=& q_D \biggl[ \frac{1}{\sum_{i=1}^{J_{t+1|t}}
w_{t+1|t}^{(i)}}
\frac{\sum_{n=m+1}^\infty
P^n_{m+1}p_{t+1|t}(n)
q_D^{n-(m+1)}}{\sum_{n=m}^\infty
P^n_{m}p_{t+1|t}(n)q_D^{n-m}} \biggr] D_{t+1|t}(x)
\nonumber
\\[-8pt]
\\[-8pt]
\nonumber
&&{}+p_D \sum_{z \in Z_{t+1}} \sum
_{i=1}^{J_{t+1|t}} \bar{w}_{t+1|t}^{(i)}(z)N
\bigl(x;\mu_{t+1}^{(i)}(z),P_{t+1}^{(i)}
\bigr),
\end{eqnarray}
where
\begin{eqnarray*}
\bar{w}_{t+1|t}^{(i)}(z)&=&
\frac{w_{t+1|t}^{(i)}q_{t+1}^{(i)}(z)}{\sum_{i=1}^{J_{t+1|t}}
w_{t+1|t}^{(i)}q_{t+1}^{(i)}(z)},\\
q_{t+1}^{(i)}(z)&=&N
 \bigl(z;H_{t+1}\mu_{t+1|t}^{(i)},R_{t+1}+H_{t+1}P_{t+1|t}^{(i)}H_{t+1}^T
 \bigr).
 \end{eqnarray*}
 The mean and the
 covariance matrix are
$\mu_{t+1}^{(i)}(z)=\mu_{t+1|t}^{(i)}+K_{t+1}^{(i)}(z-\break H_{t+1}\mu
_{t+1|t}^{(i)}),
  P_{t+1}^{(i)}=[I-K_{t+1}^{(i)}H_{t+1}]P_{t+1|t}^{(i)}$,
respectively, where
$K_{t+1}^{(i)}=\break P_{t+1|t}^{(i)}H_{t+1}^T(R_{t+1}+H_{t+1}P_{t+1|t}^{(i)}H_{t+1}^T)^{-1}$.
Furthermore, the posteriorcardinality distribution is propagated via
the following equation:
%
\begin{equation}
\label{gcorrZFAcard} p_{t+1|t+1}(n) = p_{t+1|t}(n) \frac{
P^n_{m}q_D^{n-m}}{\sum_{l=m}^\infty
P^l_{m}p_{t+1|t}(l)q_D^{l-m}}.
\end{equation}
\end{proposition}

\begin{remark}
If there were only one intracellular movement during the tracking time
and neither a birth nor a death of an organelle were allowed, then
Propositions \ref{gpred} and~\ref{gcorr} would yield the
special case of monitoring a random singleton, that is, one organelle
in our experiments. Furthermore, if the probability of detection
$p_D=1$ (thus $q_D=0$) and there was one component in the Gaussian
mixture, then equation \eqref{gcorrZFAPHD} would yield the typical
Kalman filter update equation and in this special case the matrix $K$
would play the role of the Kalman gain matrix.
\end{remark}

\section{Results} \label{results}
Having established the theoretical framework, we present our biological
data analysis and tracking in this section. We start with a summary of
our algorithm.

\textit{Step \textup{0:} Initialization}. The initial intensity, $D_{0|0}$, is
considered as a Gaussian mixture with $J_0$ components. Furthermore, the
initial cardinality distribution, $p_{0|0}(n)$, is considered a priori
to a single
object.

\textit{Step \textup{1:} Prediction}. At time $t$ the predicted intensity
$D_{t+1|t}$ is a Gaussian mixture whose components' weights, means and
covariance matrices are derived in equation (\ref{gpredZFAPHD}).
Equation (\ref{gpredZFAcard}) yields the corresponding
posterior predictive cardinality distribution, $p_{t+1|t}(n)$.

\textit{Step \textup{2:} Update}. At time $t+1$, the predictions generated in
Step 1 are updated based on new measurements. More precisely, the
posterior PHD, $D_{t+1|t+1}$, is a Gaussian mixture whose weight, mean
and covariance matrix is derived by equation (\ref{gcorrZFAPHD}). The
posterior cardinality distribution, $p_{t+1|t+1}(n)$, is estimated
according to equation (\ref{gcorrZFAcard}).

\textit{Step \textup{3:} Merging and pruning}. The number of Gaussian components
increases as time progresses. In fact, at a given time, $t$, the
Gaussian mixture will require $\mathcal{O}(J_{t-1}|Z_t|)$ components,
where $J_{t-1}$ is the number of components of the posterior intensity
$D_{t-1|t-1}$ at time $t-1$. Since components with low weight do not
provide any significant contribution to the approximation of the
posterior multi-target density, we eliminate the components whose
weights are negligible and below some preset threshold, $T$ (e.g.,
$T=10^{-5}$). The remaining components of the mixture are renormalized
such that their sum equals 1.

Furthermore, there are components which are close to each other and
practically could be approximated by a single Gaussian distribution.
Indeed, if two components of the mixture with weight, state and
covariance, $(w_i,x_i,P_i)$ and $(w_j,x_j,P_j)$, respectively, have
distance $d_{i,j} \doteq(x_i-x_j)P_i^{-1}(x_i-x_j)^t$ less than some
threshold, $U$, then these mixing components are merged into one
[\citet
{CPV}]. The threshold $U$ should be chosen much smaller (e.g., $U=0.004$)
than the standard deviation of the observations' noise so that the
filtering algorithm does not consider two different objects as one when
they are close together, such as when their paths are crossing each other.

\textit{Step \textup{4:} Multi-object state extraction}. To extract the
organelles' states, we focus on only the modes of the corresponding
Gaussian mixture. The number of
organelles is estimated from the cardinality distribution using a
maximum a posteriori (MAP)
estimator $\hat{n}= \arg\sup_n p(n|Z_{1:t})$.

Schematically, the algorithm works in the following way, for all
$t=0,1,\ldots$:
\[
(D_{t|t},p_{t|t}) \stackrel{\mathrm{Proposition~\scriptsize{\ref{gpred}}}}
{\longrightarrow} (D_{t+1|t},p_{t+1|t})\stackrel{\mathrm{Proposition~\scriptsize{\ref{gcorr}}}} {\longrightarrow} (D_{t+1|t+1},p_{t+1|t+1}),
\]

where the PHD, $D_{\cdot|\cdot}$, is estimated via the triplet of
weights, mean and covariance.
%
%
\subsection{Synthetic data} \label{synthetic data}
This section illustrates a simulated scenario with respect to organelle
movements. Consider a set $X_t=\{ x_t^1, x_t^2, \ldots, x_t^{n_t}\} $
whose members are 4-dimensional state vectors of the $n_t$ organelles
at time $t$. Precisely, an organelle's state vector is $x_t^i
\doteq[p_{x,t}, v_{x,t}, p_{y,t},v_{y,t}]^T$ for any $i=1,\ldots,n_t$,
where $(p_{x,t}, p_{y,t})$ denote the spatial coordinates of the
organelle on the $xy$-plane and the corresponding velocities are
denoted as
$(v_{x,t},v_{y,t})$. The movements in a cell may be considered to take
place in a force field\vadjust{\goodbreak} which is on average inactive. However, when a
molecular motor exerts a pushing force on an organelle, then there is a
positive deviation from the mean zero. By the same token, when friction
and/or other large enough backward-acting forces occur, then the
organelles will slow down and eventually stop, and thus a symmetric
negative deviation from the mean-zero force field is caused. Therefore,
one may consider that the force field is normally distributed with mean
zero and pertinent covariance. This consideration is actually validated
in Section~\ref{data} where experimental data are analyzed. Given that
the mass is conservative over time frames considered here (a valid
assumption), Newton's second law yields that the acceleration, $a$,
follows a normal distribution. The velocity changes in turn are also
normally distributed where the covariance depends on the size of the
time intervals. Given the fact that $\dot{p}_{x,t}=v_{x,t}$ and $\dot
{p}_{y,t}=v_{y,t}$, we can formally state the following linear
stochastic differential equations system:
%
\begin{equation}
\label{ctssde} \pmatrix{ dp_{x,t}
\vspace*{2pt}\cr
dv_{x,t}
\vspace*{2pt}\cr
dp_{y,t}
\vspace*{2pt}\cr
dv_{y,t} }= \pmatrix{ v_{x,t}
\vspace*{2pt}\cr
0
\vspace*{2pt}\cr
v_{y,t}
\vspace*{2pt}\cr
0 } \,dt+ \pmatrix{ 0&0
\vspace*{2pt}\cr
\sigma_x & 0
\vspace*{2pt}\cr
0&0
\vspace*{2pt}\cr
0& \sigma_y } \pmatrix{ du_{x,t}
\vspace*{2pt}\cr
du_{y,t} },
\end{equation}
where $u_{x,t}, u_{y,t}$ are independent Brownian motions and the
driving noises $\sigma_x \dot{u}_{x,t} $ and $\sigma_y \dot
{u}_{y,t} $
are Gaussian noises with covariances $\sigma_x^2 \delta(t)$ and
$\sigma
_y^2 \delta(t)$, respectively, where $\delta(t)$ is the delta function.
Discretizing and approximating the system \eqref{ctssde}, we have a
two-dimensional model given below:
%
\begin{equation}
\label{2dconvel} \pmatrix{ p_{x,t}
\vspace*{2pt}\cr
v_{x,t}
\vspace*{2pt}\cr
p_{y,t}
\vspace*{2pt}\cr
v_{y,t} }=\pmatrix{ 1 & \Delta& 0 & 0
\vspace*{2pt}\cr
0 & 1 & 0 & 0
\vspace*{2pt}\cr
0 & 0 & 1 & \Delta
\vspace*{2pt}\cr
0 & 0 & 0 & 1 } \pmatrix{ p_{x,t-1}
\vspace*{2pt}\cr
v_{x,t-1}
\vspace*{2pt}\cr
p_{y,t-1}
\vspace*{2pt}\cr
v_{y,t-1} } + \pmatrix{ \displaystyle\frac{\Delta^2}{2}&0
\vspace*{2pt}\cr
\Delta& 0
\vspace*{2pt}\cr
0&\displaystyle\frac{\Delta^2}{2}
\vspace*{2pt}\cr
0& \Delta }\xi_{t-1},
\end{equation}
where the model noise, $\xi_{t-1}$, is a collection of independent
Gaussian random variables with covariance matrix $\Sigma= \diag\{
\sigma
_x^2,\sigma_y^2\}$.
The sampling time is considered $\Delta=1$~s since data from organelles'
movements are collected every one second. The velocity changes are
normally distributed with mean zero, and thus 99.7\% of the data are
within three standard deviations from zero. Taking into consideration
the biological finding that organelles may move up to 7~$\mu$m$/$s (in
both directions) [\citet{TYO}], the standard deviation coefficients are
chosen $\sigma_x=\sigma_y=2.33$~$\mu\mbox{m}/ \mbox{s}^2$.
If one decreased or increased drastically the variance, then the
estimates would not be accurate. Small noise dynamics (e.g., $\sigma
_x=\sigma_y=0.1$~$\mu\mbox{m}/ \mbox{s}^2$) yield predictions based
on almost perfect linear dynamics which could lead to erroneous
estimation in case organelles exhibit a slightly curvy behavior. By the
same token, a large standard deviation (e.g., $\sigma_x=\sigma_y=5$~$\mu
\mbox{m}/ \mbox{s}^2$) produces a wide range of samples which lead to
inaccurate estimates.

Furthermore, each object is considered with survival probability,
$p_{S,t}=0.99$, such that any organelle within the tracking domain is
under monitoring unless its signal disappears. The maximum number of
involved Gaussian components is considered to be fairly large,
$N_{\mathrm{max}}=200$. The object-birth process is a Poisson RFS with intensity
defined as in (\ref{bithgauss}), where $w_b=0.25$, $\mu_b^{(1)}=[3\ 0\ 5\  0]^T$, $\mu_b^{(2)}=[4\  0\   -\!6\   0]^T, \mu_b^{(3)}=[-3\   0\
-\!2\ 0]^T$, $\mu_b^{(4)}=[-4\   0\   8\   0]^T$, and $P_b=10\mathbb{I}_4$.
The four different means, $\mu_b^{(i)},  i=1,\ldots, 4$ are selected
to ensure that births on all four quadrants are considered with equal
probability $w_b=0.25$. The covariance of the birth intensity is also
large such that a vast candidate area of newborn organelles is covered.
Given that our experimental environment did not suffer from low
signal-to-noise ratio and no false alarm occurred, the
probability of detecting an organelle is state independent and equals
$p_{D,t}=0.98$.

We first focus on the synthetic data which consist of the spatial
coordinates. Consider at given time $t+1$ the random set, $Z_{t+1}=\{
z_{t+1}^1, z_{t+1}^2, \ldots,\break  z_{t+1}^{m_{t+1}}\}$, where for each
$i$ the data $z^i_{t+1}=(p_{x,t+1},p_{y,t+1}), i=1,\ldots, m_t$, is a
two-dimensional vector whose likelihood is defined in (\ref
{gausslik}), with
%
\begin{equation}
\label{likdet} H_t=\pmatrix{ 1 & 0 & 0 & 0
\vspace*{2pt}\cr
0 & 0 & 1 & 0},\qquad R_t= \sigma^2_o
\mathbb{I}_2,
\end{equation}
and $\sigma_o=0.2$~$\mu$m is the standard deviation of the
measurement noise due to optical limitations and experimental noise.
For example, there is a fundamental maximum to the resolution of any
optical system due to diffraction. The diffraction defines the
microscope's point-spread function which describes the response of an
imaging system to a point light source. Furthermore, our procedure uses
a weight threshold
$T=10^{-5}$ for the pruning procedure and a threshold $U=0.004$ for
the merging part of the algorithm (step 3 in the algorithm).

The synthesized organelles' trajectories, which play the role of the
true trajectories, are created by evolving a number of organelles
according to dynamics \eqref{2dconvel}, and the corresponding
observations were created after perturbing the true trajectories by a
normally distributed noise with covariance $R_t$ as in \eqref{likdet}.

\begin{figure}

\includegraphics{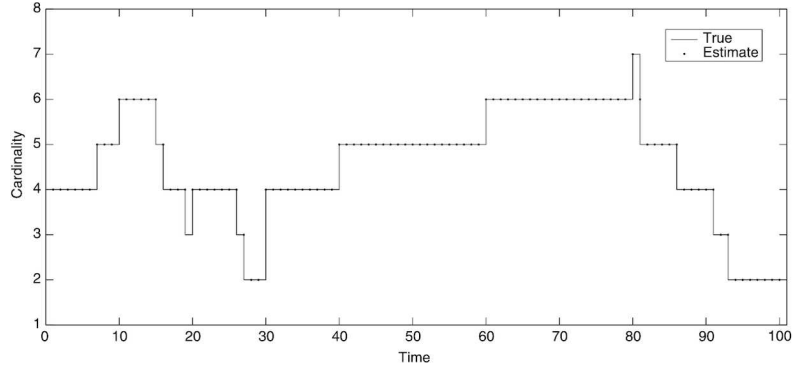}

\caption{Number of organelles per time step.}
\label{cardsim}
\end{figure}

Figure~\ref{cardsim} shows that there are twelve organelles (in total)
which are monitored for 100 time steps. At any given time $t$, the
number of organelles is unknown a priori and is not fixed, that is,
random birth and death of organelles are allowed with pertinent
dynamics based on Assumption~\ref{gbd}. In fact, the organelles' number
increases and decreases drastically during the first thirty steps and
the last twenty ones as well. This makes the problem a rather
formidable one by keeping in mind that previous studies have monitored
simultaneously a fixed and a priori known number of intracellular
movements with overall known dynamics, for example, \citet{SNM}. In
contrast, our algorithm assumes an initial cardinality of 1 (see step 1
of the algorithmic description) and updates its estimate based on
available data. Thus, our algorithm captures accurately all
modifications in the number of organelles and it gives an accurate estimate.
%

%
\begin{figure}[b]

\includegraphics{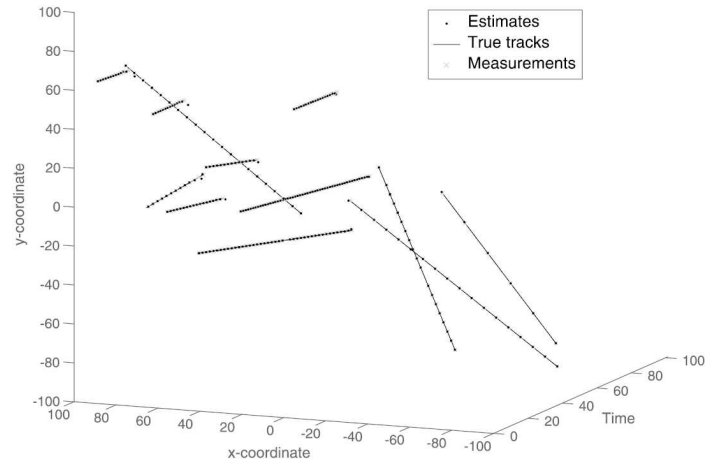}

\caption{Linear trajectories of organelles in the $xy$-plane over time.}
\label{xysims}
\end{figure}

Figure~\ref{xysims} shows a three-dimensional graph of the
trajectories' estimates of the organelles across time. As we can see,
there are several crossings, often in the $y$-direction. Tracking
methods for intracellular movements that assume one-to-one
correspondence between a measurement and an object fail to resolve the
most ambiguous track interaction scenarios, for example, when objects
are in close proximity. However, in our case, we do not assume\vadjust{\goodbreak} any sort
of prior one-to-one correspondence, instead we employ a multi-object
statistical framework by considering a single set of objects, thereby
producing accurate estimates even during the difficult occasions such
as crossings.

Indeed, the estimates are very close to the true trajectories, but to
quantify any sort of error a multi-object error distance is considered.
The characteristics of a multi-object distance should (1) be a metric
on the space of finite sets, (2)~capture cardinality and state errors
and (3) have a physical interpretation. Toward this end, we employ a
metric from point processes theory in order to measure the discrepancy
between the estimates and the true values [\citet{Bre,MoWa}]. A formal
definition of this metric according to \citet{SVV} is given below.

\begin{definition}\label{ospadefn}
Let $W \subset\mathbb{R}^N$ be a closed and bounded observation window
and $d$ denote the Euclidean metric. For $c > 0$, let $d^{(c)} (x, y)
\doteq\min(c,\break d(x, y))$ denote the distance between $x,y \in W$ and
$P_n$ denote the set of permutations on $\{1,2,\ldots,n\}$ for any $n
\in\mathbb{N}$. For $1 \leq\ell< \infty,  c > 0$ and arbitrary
finite subsets $X = \{x_1,\ldots,x_m\}$ and $Y = \{y_1,\ldots,y_n\}$ of
$W$, where $m,n=0,1,2,\ldots,$ define for $m \leq n$,
%
\begin{equation}
\label{ospa} \bar{d}_\ell^{(c)}(X,Y) \doteq \Biggl(
\frac{1}{n} \Biggl( \min_{\pi
\in
P_n} \sum
_{i=1}^m d^{(c)}(x_i,y_{\pi(i)})^\ell+c^\ell(n-m)
\Biggr) \Biggr)^{1/\ell},
\end{equation}
and $\bar{d}_\ell^{(c)}(X,Y)=\bar{d}_\ell^{(c)}(Y,X)$ if $m>n$.
Moreover, if $\ell=\infty$, then
%
\begin{eqnarray}
\bar{d}_\infty^{(c)}(X,Y) &=&\min_{\pi\in P_n} \max
_{1\leq i \leq n} d^{(c)}(x_i,y_{\pi(i)})\qquad
\mbox{if } m=n
\nonumber
\\[-8pt]
\\[-8pt]
\nonumber
&= & c\qquad \mbox{if } m \neq n.
\end{eqnarray}
For any $\ell\in[1,\infty]$ the distance is equal to zero if $m = n =
0$. The function $\bar{d}_\ell^{(c)}(X,Y)$ is called the Optimal
SubPattern Assignment (OSPA) metric of order $\ell$ with cutoff
parameter $c$.
\end{definition}

%
\begin{remark}
\citet{ScXi} examined the special case for $\ell=c=1$ and \citet{SVV}
generalized it for any $\ell,c$. The metric $\bar{d}_\ell^{(c)}$ is
based on a Wasserstein construction. The advantage of this metric is
that equation \eqref{ospa} takes into consideration the error due to
localization and cardinality at the same time. An alternative measure
of discrepancy is the Haussdorff distance [\citet{MoWa}], however, it
is relatively insensitive to difference in cardinality as was noted in
\citet{HoMa}. The order parameter $\ell$ is similar to the parameter of
the $\ell$th order Wasserstein metric between the empirical
distributions of the point patterns $X$ and $Y$. Furthermore, given
that $c$ is fixed, the parameter $\ell$ in \eqref{ospa} assigns more
weight to outliers. The metric $\bar{d}_\ell^{(c)}(X,Y) \in[0,c]$ for
any $c>0$ in turn gives us a measure of performance with respect to the
worst possible distance $\ell$. Also, if $0<c_1<c_2<\infty$, then
$\bar
{d}_\ell^{(c_1)} \leq\bar{d}_\ell^{(c_2)}$. Moreover, the cutoff
parameter $c$ determines the weighting of how the metric penalizes
cardinality errors as opposed to localization errors. Smaller values of
$c$ tend to put emphasis on the localization error and make the metric
unchanged by cardinality errors. Thus, the designer can determine how
strongly a false or missing estimate will be penalized by modifying the
value of $c$. Here, we have chosen $\ell=1$ and $c=30$ such that the
OSPA is sensitive enough in both localization and cardinality errors.
The choice of the value $\ell=1$ has the benefit that the OSPA-metric
measures a first order per-object error and that the sum of
localization and cardinality components equals the total metric. The
reader should refer to \citet{SVV} and the references therein for
further details on the OSPA metric.\looseness=1
\end{remark}

\begin{figure}[b]

\includegraphics{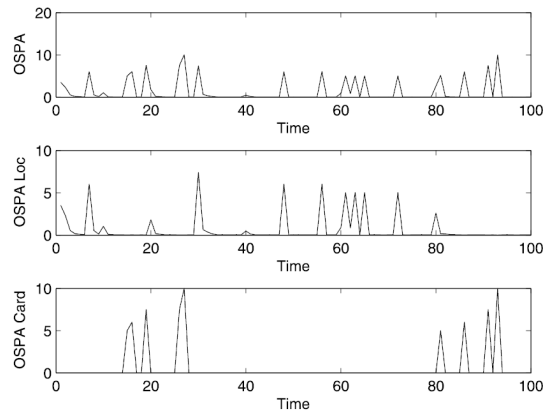}

\caption{Error measured via the OSPA metric. The error can be as large
as the cutoff parameter $c=30$.}
\label{ospasims}
\end{figure}
The top picture in Figure~\ref{ospasims} depicts the error using the
OSPA metric given in equation \eqref{ospa}. We observe that large
errors (peaks in the figure) occur when the organelles are crossing and
when there is a change of the number of organelles (e.g., at $t=20$).
This is expected since these are the most difficult situations. The
OSPA error cannot exceed the value 30 since the cutoff parameter is set
at $c=30$, however, even in the most difficult cases, the error remains
well below 10. The two subsequent pictures are showing localization and
cardinality error. The localization errors for two patterns
$X=(x_1,\ldots,x_m)$ and $Y=(y_1,\ldots,y_n)$ with $m \leq n$ and
$\ell
< \infty$ are given by\looseness=1
\begin{eqnarray*}
\bar{e}_{\ell,\mathrm{loc}}^{(c)}(X,Y) &=& \Biggl(\frac{1}{n} \Biggl(
\min_{\pi
\in
P_n} \sum_{i=1}^m
d^{(c)}(x_i,y_{\pi(i)})^\ell \Biggr)
\Biggr)^{1/\ell},\\
\bar{e}_{\ell,\mathrm{card}}^{(c)}(X,Y)&=& \biggl(
\frac{c^\ell(n-m)}{n} \biggr)^{1/\ell}. %
\end{eqnarray*}\looseness=0%
Strictly speaking, the two errors, $\bar{e}_{\ell,\mathrm{loc}}^{(c)}$ and
$\bar
{e}_{\ell,\mathrm{card}}^{(c)}$, are not metrics on the space of finite subsets,
but one may still gain some insight about the performance of the filter
[\citet{SVV}].

\subsection{Experimental data} \label{data}
\begin{figure}[b]

\includegraphics{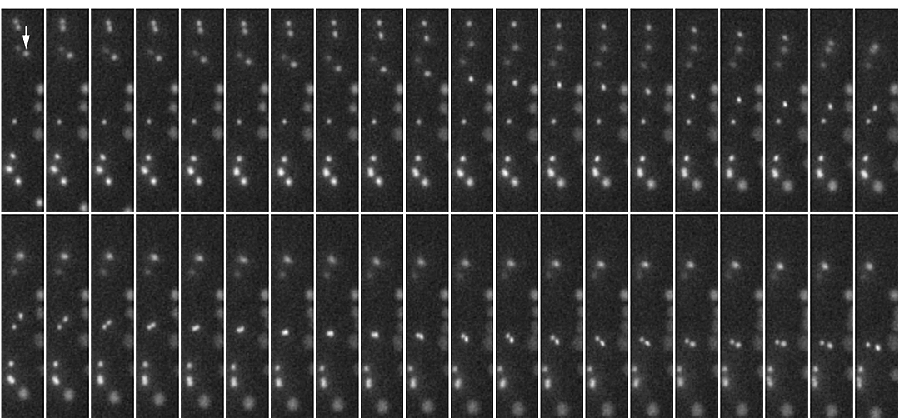}

\caption{Peroxisomes movements.}
\label{realpics}
\end{figure}

Before outlining our results, we will briefly describe the conditions
under which the movement data were retrieved. Organelles were labeled
with fluorescent protein fusions in root cells of the model plant
\textit{Arabidopsis thaliana} and cells on the surface of roots were
observed on a fluorescent microscope as described in \citet{NCN}.
Images were taken with a digital camera at regular intervals (1~s) to
generate time-lapse sequences of 1 to 2 minute duration (i.e., 60 to
120 images). These image sequences (e.g., Figure~\ref{realpics})
displayed bright spots of different sizes and intensities depending on
the size and position of the organelle relative to the focal plane.
Movements of individual organelles were readily apparent by comparing
the changes in position of spots between image frames (arrow in
Figure~\ref{realpics}). Specifically, Figure~\ref{realpics} shows the
movement of peroxisomes, small spherical organelles involved in
detoxification of reactive oxygen species which have recently emerged
as important regulators of plant growth and stress responses [\citet
{KlHe}]. Similar movements can also be observed for other organelles,
such as Golgi stacks [\citet{Neb}] and mitochondria [\citet{GKV}].

%
%

Images were analyzed quantitatively by manually marking the center of
each spot in every frame of the time-lapse sequence which was then
recorded by the Manual Tracking plugin in ImageJ [\citet{SRE}]. This
procedure produced a series of $(x,y)$ coordinates per image frame that
were manually linked to specific $(x,y)$ coordinates in subsequent
frames. The procedure of manually linking is typically slow (about 1
hour for the data set that is analyzed herein) and bias due to human
decision in linking can be a frequent disadvantage. In our case, the
resulting two-dimensional vectors declaring the position of organelles
on the $xy$-plane at every time point were used (1) to calculate the
instantaneous velocities of the organelles over time; (2) to provide
experimental values for the accelerations' distributions; and (3) to
provide the raw data to the statistical tracking algorithm without
knowing a priori which data (coordinates) correspond to which organelle.

\begin{table}[b] 
\tablewidth=150pt
\caption{$p$-values of two Kolmogorov--Smirnov tests for the
acceleration data points of peroxisomes}
\label{normtest}
\begin{tabular*}{150pt}{@{\extracolsep{\fill}}lcc@{}}
\hline
\textbf{Acceleration} & $\bolds{p}$\textbf{-value} & $\bolds{H_0}$\\
\hline
$a_x$ & 0.31\phantom{00} & Accept \\
$a_y$ & 0.3265 & Accept \\
\hline
\end{tabular*}
\end{table}

\begin{figure}[t]

\includegraphics{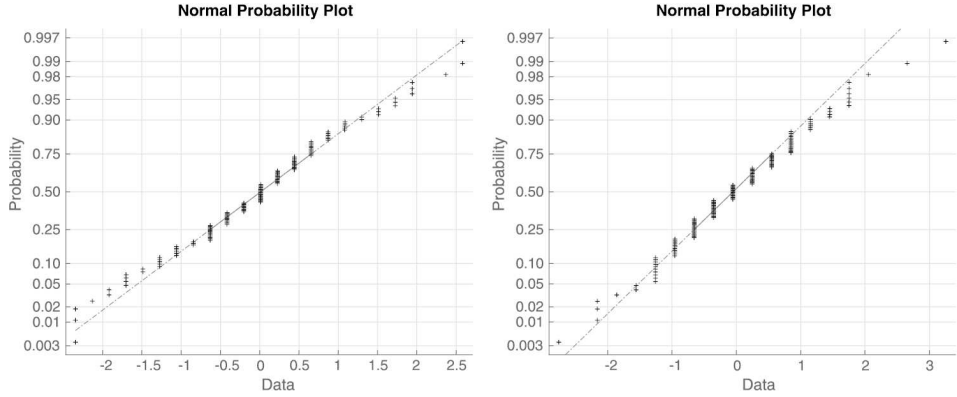}

\caption{Testing normality of the organelles' acceleration. \emph
{Left panel}:
Acceleration on the $x$-axis. \emph{Right panel}: Acceleration on the
$y$-axis.}\label{normxy}
\end{figure}

In the following, we focus on the motions of eight peroxisomes
retrieved in experiments in the second author's lab. First, we
decompose the acceleration, and we investigate the distributional
behavior of the accelerations per coordinate separately based on the
experimental data. There are $m=284$ acceleration data points from the
eight peroxisomes with mean and standard deviation on the $x$-axis,
$\mu
_x^a=-0.0326,   \sigma_x^a=0.9998$, respectively. The corresponding
mean and standard deviation on the $y$-axis are $\mu_y^a=0.0429,
\sigma_y^a=0.6922$. Next, we test if the accelerations follow a normal
distribution using a Kolmogorov--Smirnov test and visually by plotting
two normality plots, one per coordinate. As we can see from the results
of the Kolmogorov--Smirnov tests presented in Table~\ref{normtest}, and
the normal probability plots in Figure~\ref{normxy}, the two
accelerations of the eight peroxisomes follow a Gaussian distribution.
Thus, the arguments of Section~\ref{synthetic data} imply that the
dynamics of the eight peroxisomes can be described by the discrete
system in \eqref{2dconvel}.

%
\begin{figure}[b]

\includegraphics{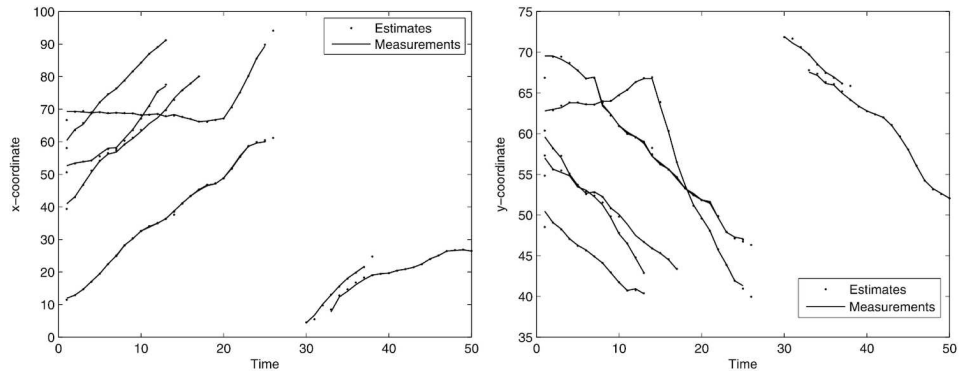}

\caption{Trajectories of organelles. \emph{Left panel}: Trajectories in
the $x$-direction. \emph{Right panel}: Trajectories in the $y$-direction.}
\label{xpaths}
\end{figure}

\begin{figure}[t]

\includegraphics{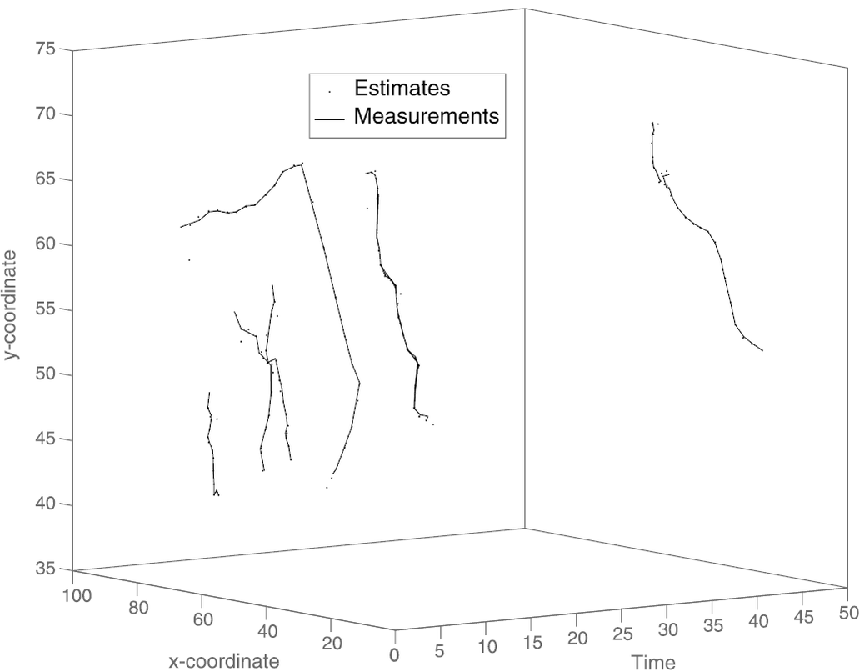}

\caption{Trajectories of organelles in the $xy$-plane over time.}
\label{xypaths}
\end{figure}

\begin{figure}[b]

\includegraphics{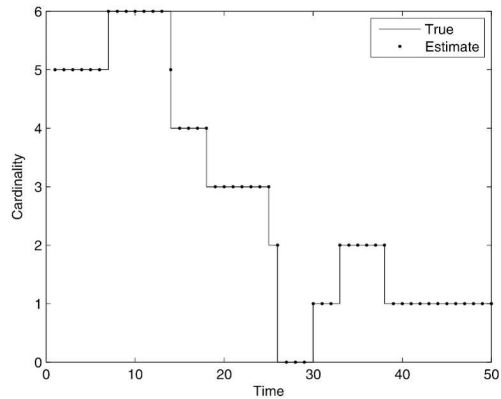}

\caption{Number of organelles per time step.}
\label{card}
\end{figure}

Therefore, employing the dynamics \eqref{2dconvel} accompanied by the
several hyperparameters discussed in Section~\ref{synthetic data}, we
describe our findings for the motions of the peroxisomes. Figures~\ref
{xpaths} and~\ref{xypaths} show the trajectories based on
measurements (line) and the corresponding estimates represented as dots
in the figures. At the initial time step, Figure~\ref{xpaths} shows a
greater mismatch between the estimates and the data than in the next
sampling periods. This is expected since the algorithm attempts to
``learn'' the pattern of the organelles' motion. Although the
peroxisomes' overall trajectories are not linear, they are piecewise
linear per time step~(1~s), and thus the dynamics of Section~\ref
{synthetic data} perform satisfactorily since sampling occurs every
$\Delta=1$~s. If the piecewise linearity was violated and/or the
acceleration distribution was heavy tailed, then the dynamics in
equation \eqref{2dconvel} would produce errors which would depend on
the curvature of the true trajectories and/or the non-Gaussian noise.
Figure~\ref{card} depicts the cardinality (number of peroxisomes) per
time step. As we observe, the CPHD filter accurately captures the
target number when their number does not vary, and it takes 1 to 2
sampling time steps to realize the change in the organelle number.
Also, the algorithm correctly estimates that there were not any
organelles to monitor during the time interval $[26,29]$. The duration
of the automated tracking process based on our algorithm is about 10~s
versus roughly 1~hr for the manual tracking of the same eight peroxisomes.

Due to lacking the true trajectories of the organelles (in fact, it is
impossible to know them with the current technology) [\citet{SNM}], the
OSPA measurement of error (and any other metric of this type) cannot be
used since it measures the discrepancy between the algorithmic
estimates and the true trajectories (not the observed measurements).
However, according to our simulation results exposed in Section~\ref
{synthetic data}, we believe that our estimates are very close to the
true trajectories of the eight peroxisomes.
%
%
\section{Summary and discussion} \label{summary}
In this paper we have considered the motion of organelles as evolving
sets. This succeeded by incorporating random sets techniques for
multi-object tracking and using the cardinalized probability hypothesis
density filter. Employing a novel Gaussian mixture implementation of
the CPHD filter, we were able to successfully generate an automated
method for a quantitative analysis of intracellular movements, which
took about 10 seconds versus about 1 hour for manually linking the same
data. The new approach's computational cost was linearly dependent on
the number of objects multiplied by the number of data points. Our
model was capable of simultaneously monitoring a large number of
organelles, specifically peroxisomes, and distinguishing them even when
they were in close proximity. Consequently, not only did our algorithm
monitor the organelles but it also gave an accurate estimate on the
number of organelles without assuming a fixed and known number of them.
Furthermore, our data analysis revealed that the acceleration of the
peroxisomes are mean-zero normally distributed, which according to
Newton's second law supports an on average ``inactive'' force field
within a cell where positive (pushing force by the myosin motors) or
backward-acting forces (e.g., friction) are developed in a symmetric
fashion given that mass is conservative. Consequently, the two
parameters, myosin power and local friction, were fairly constant on
average over time and space, respectively. On the other hand, large
changes in velocity (if any) presumably would result from a static
organelle engaging with a cytoskeletal track, or from a moving
organelle dropping from a cytoskeletal track. We expect these changes
to occur nearly instantaneously, however, technical limitations
prevented us from detecting these very rapid changes if they indeed
existed. In particular, we had to employ exposure times up to 100 ms to
obtain sufficient signal for organelle detection. In addition, images
were taken in 1~s intervals and had a nominal resolution of 200 nm per
pixel. Given that myosin motors take 35 nm steps and can move up to 7~$\mu$m$/$s,
that is, one step every 5 ms, as noted in \citet{TYO}, it
is apparent that these imaging parameters do not allow us to capture
the anticipated very fast acceleration and deceleration events
directly. Instead we can only compute the integrated behavior of
organelles over many individual myosin steps. Therefore, this
scientific conjecture regarding changes in organelle velocities should
be further examined on large experimental data sets which could yield a
more detailed distribution of accelerations, dynamics and thus
potentially the mechanics within a cell overall.

Focusing on the algorithm itself, although it captures the organelles'
behavior accurately, it did not take other scenarios into consideration
which would increase the already severe complexity of the problem. For
example, there might be cases where organelles may move in a more
erratic fashion. In this scenario, the acceleration distribution might
not be normally distributed and thus nonlinear and/or nonGaussian
dynamics could be fruitful for such data. A possible future research
avenue is to use high noise with suitably controlled drift dynamics or
a more complex autoregressive model. Another way is to approximate the
overall nonlinearities and/or add more experimental features, for
example, include information about the shape and signal intensity of
organelles in the linking step [\citet{SbKo,SDG,SNM,Sma}]. Moreover,
the organelles' survival and detection probabilities were presumed
state independent and time invariant. On the other hand, these
probabilities clearly depend on the position of organelles in a cell.
For instance, organelles in close proximity to each other may not be
detected or, given the curvature of cells, the survival probability of
an organelle will decrease as it approaches an out-of-focus region of
the cell. In our experimental data, crossings occurred only a few times
and organelles were always in-focus and ``disappeared'' when they exited
the focal domain. Attempting to bypass Assumption~\ref{prob},
techniques developed in \citet{HFH,HuFr} may be fruitful for these
difficult scenarios.

In conclusion, this manuscript offers the establishment of a systematic
way of creating an automated algorithm for monitoring motility within a
cell by considering a unifying statistical framework for multiple
objects. In turn, such an automated tracking algorithm will greatly
strengthen the study of motion patterns in cells. Consequently,
understanding the typical behavior of healthy molecular processes will
have a great impact in quickly recognizing abnormalities associated
with disorders.

\section*{Acknowledgments}
The authors would like to thank the Editor, Professor Karen Kafadar, an
anonymous Associate Editor and two anonymous reviewers for their
comments which allowed us to substantially improve our manuscript. The
first author would like to thank Dr. Mahler for introducing him into
this fascinating topic of statistical research and for fruitful
discussions. Part of this research was established while the authors
collaborated in a NIMBioS Investigative Workshop.

%

\printaddresses
\end{document}